\documentclass[12pt,a4paper]{article}
\usepackage{amsmath}
\usepackage{epsfig}

\input{epsf}

\newcommand{\mc}{\multicolumn}
\newcommand{\eref}[1]{(\ref{#1})}

\newcommand{\rf}[1]{(\ref{#1})}
\newcommand{\beq}{\begin{equation}}
\newcommand{\eeq}{\end{equation}}
\newcommand{\bea}{\begin{eqnarray}}
\newcommand{\eea}{\end{eqnarray}}

\newcommand{\e}{\mathrm{e}}

\renewcommand{\L}{\Lambda}
\renewcommand{\b}{\beta}
\renewcommand{\a}{\alpha}
\newcommand{\n}{\nu}
\newcommand{\m}{\mu}

\newcommand{\ep}{\varepsilon}

\newcommand{\del}{\delta}

\newcommand{\oh}{\frac{1}{2}}
\newcommand{\oq}{\frac{1}{4}}

\newcommand{\ra}{\right\rangle}
\newcommand{\la}{\left\langle}

\newcommand{\cD}{{\cal D}}

\newcommand{\cM}{{\cal M}}

\begin{document}
\topmargin 0pt
\oddsidemargin 5mm
\headheight 0pt
\headsep 0pt
\topskip 9mm

\hfill    NBI-HE-97-20\\

\hfill    DEMO-HEP-97-10

\begin{center}
\vspace{24pt}
{ \large \bf Spikes in  Quantum Regge Calculus}

\vspace{24pt}

{\sl Jan Ambj\o rn, Jakob L.\ Nielsen, Juri Rolf}

\vspace{12pt}
The Niels Bohr Institute, University of Copenhagen, \\
Blegdamsvej 17, DK-2100 Copenhagen \O , Denmark

\vspace{24pt}
 and
\vspace{24pt}

{\sl  George Savvidy}

\vspace{12pt}

National Research Center ``Demokritos'', \\
Ag. Paraskevi, GR-15310, Athens, Greece

\vspace{36pt}

\end{center}

\vfill

\begin{center}
{\bf Abstract}
\end{center}

\vspace{12pt}

\noindent
We demonstrate by explicit calculation of the DeWitt-like 
measure in two-dimensional quantum Regge gravity 
that it is highly non-local and 
that the average values of link lengths $l$, $<l^n>$,
do not exist for sufficient high powers of $n$. Thus  
the concept of length has no natural definition
in this formalism and a generic manifold degenerates into spikes. This might
explain the failure of quantum Regge calculus to reproduce
the continuum results of two-dimensional quantum gravity.
It points to severe problems for the Regge approach  in higher dimensions.

\vfill

\newpage

\section{Introduction}
\label{sec:intro}
In the path integral formulation of quantum gravity we are instructed 
to integrate over all gauge invariant classes of metrics on a 
given manifold $\cM$:
\begin{equation}
Z(\L,G) = \int  \cD [g_{\m\n}] \; {\e^{-S_{EH}[g_{\m\n}]}},
\label{*1}
\end{equation}
where $[g_{\m\n}]$ denotes an equivalence class of metrics, i.e.\
the metrics related by a diffeomorphism $\cM \mapsto \cM$, and 
 $S_{EH}[g_{\m\n}]$ denotes the Einstein-Hilbert action:
\begin{equation}
S_{EH}[g_{\m\n}]= \int_{\cM} \sqrt{g} \; \Bigl(\L -\frac{1}{16 \pi G} R\Bigr). 
\end{equation} 
We have written the functional integral in Euclidean space-time. In 
two-dimensional space-time this integral is well defined. In 
higher dimensions special care has to be taken, either by adding 
higher derivative terms  
to the Einstein-Hilbert action or by performing 
special analytic continuations of the modes which in higher dimensions
are responsible for the unboundedness of the Einstein-Hilbert action.
In this paper we will mainly discuss two-dimensional gravity,
but as will be clear the simple underlying reason for the 
problems we encounter are likely to persist in higher dimensions.

It is well know how to perform the functional integration
\rf{*1}. One introduces on the space of all metrics of a given manifold
$\cM$ the norm
\begin{equation}
  \label{line-element}
  ||\del g_{\m\n} ||^2 =\int_{\cal M} d^Dx\sqrt{g}\,\delta
  g_{\mu\nu}(x)G^{\mu\nu\alpha\beta}\delta g_{\alpha\beta}(x)\, ,
\end{equation}
where $G^{\mu\nu\alpha\beta}$ is the DeWitt metric  given by
\begin{equation}
  \label{DeWitt}
  G^{\mu\nu\alpha\beta}=\frac{1}{2}(g^{\mu\alpha}g^{\nu\beta}+g^{\mu\beta}g^{\nu\alpha}
  +Cg^{\mu\nu}g^{\alpha\beta})\, .
\end{equation}
This metric can be derived from symmetry considerations and from the
requirement of diffeomorphism invariance. The constant
$C$ determines the signature of the metric. Let $D$ denote 
the dimension of space-time. For $C>-{2}/{D}$, the
metric $G^{\mu\nu\alpha\beta}$ is positive definite, and for
$C<-{2}/{D}$, it has negative eigenvalues.  In canonical quantum
gravity $C$ equals $-2$.  Using the DeWitt metric, the
functional measure on the space of all metrics can then be determined
to be:
\begin{eqnarray}
  \label{measure}
  {\cal D}g_{\mu\nu} & =&
  \prod_x\sqrt{\det(\sqrt{g}\,G^{\mu\nu\alpha\beta})}\prod_{\mu\leq\nu}
  dg_{\mu\nu}(x) \nonumber \\
  & \sim & \prod_x
  \sqrt{1+\frac{DC}{2}}\,g(x)^{\frac{1}{8}(D+1)(D-4)}
  \prod_{\mu\leq\nu}dg_{\mu\nu}(x)\,.
\end{eqnarray}
This is the unique ultra-local diffeomorphism-invariant functional
integration measure for $g_{\m\n}$. 
We get the measure $\cD [g_{\m\n}]$ in  \rf{*1} 
by an appropriate gauge fixing such that we divide \rf{measure} by the 
volume of the diffeomorphism group. 

This program has been carried out explicitly for two-dimensional gravity
by Knizhnik, Polyakov and Zamolodchikov and by David, Distler and Kawai
\cite{kpz}. One finds
\begin{equation}\label{genus}
Z(\L,G) \sim \L^{\frac{5h-5}{2}} \e^{\frac{1-h}{2G}},
\end{equation}
where $h$ denotes the number of handles of $\cM$. The continuum derivation is 
usually based on Ward identities or a self-consistent bootstrap
ansatz, although attempts of a derivation from first principle
exist. However, {\em Dynamical Triangulations} (DT) represents
a  regularization of the path integral, where an explicit 
cut-off (the link length $\ep$ of each equilateral triangle) 
is introduced \cite{dt}. Each triangulation is viewed as representing an 
equivalence class of metrics
since the triangulation uniquely defines the distances in the manifolds 
via the length assignment of the links ($\ep$ for all links) and the
assumption of a flat metric in the interior of each triangle. The 
functional integral \rf{*1} is then  replaced by the summation over 
all triangulations of a given topology, and the class of such 
metrics can be viewed as a grid in the class of
all metrics. In the limit  $\ep \to 0$ the summation over this 
grid reproduces the continuum result \rf{genus}.

An alternative and conceptually quite nice approach was 
studied in \cite{menotti}. In these papers the 
continuum functional integral over all metrics is replaced
by the continuum functional integral over so-called piecewise 
linear metrics, i.e. metrics which represent geometries 
which are flat except in a finite number of points where 
curvature is located. If one restricts the number of singularities
to be $V$, say, one can repeat the usual continuum calculations 
except that the final functional integration over the Liouville field
becomes a finite dimensional $(3V\!\!-\!6)$--dimensional integral 
(if the manifold has spherical topology). It is believed that this integral 
converges (in some suitable sense) to the continuum value
\rf{genus} for $V \to \infty$, although this has strictly speaking 
not been proven. The final, $(3V\!\!-\! 6)$--dimensional integral requires
some explicit regularization and it has not yet been possible 
to perform this integration. However, since the measure used
is the restriction of full continuum measure to the class of piecewise 
linear metrics, and since this class of metrics includes in 
particular all metrics used in DT it {\em should} give the correct answer.

In the rest of this article we will discuss a formulation of 
two-dimensional quantum gravity,  which we denote as
{\em Quantum Regge Calculus}, (QRC), and which has  been used 
extensively as a regularization of quantum gravity (for some reviews
and extensive references see \cite{hamber}).

Classical Regge Calculus is a coordinate independent discretization of
General Relativity. By choosing a fixed (suitable) triangulation
and by viewing the link lengths as dynamical variables 
one can approximate certain aspects of smooth manifolds.
Each choice of link lengths consistent with all triangle 
inequalities creates a metric assignment to the manifold if we view 
the triangulation as flat in the interior of the simplexes.
Not all such assignments correspond to inequivalent metrics, as is clear 
by considering  triangulations of two-dimensional flat space.
If we divide out this additional invariance we obtain 
for a given triangulation with $L$ links a 
$L$-dimensional subspace of the infinite dimensional space
of equivalence classes of metrics. This counting is in agreement 
with the one above, since the relation between the number of links $L$ and the 
number of vertices $V$ where the curvature is located, is given by 
$L = 3V-6$ in the case of spherical topology.
Quantum Regge Calculus is defined by performing the functional integral 
\rf{*1} on  this finite dimensional 
space and then taking the limit $L \to \infty$. 
It is the hope that this (seemingly well defined) procedure, when 
applied to the calculation of expectation values of observables,
will produce results which converge in some suitable way to the 
continuum values of the observables when $L \to \infty$.

At this point an important difference between two-dimensional 
QRC and two-dimensional DT
emerges: the first procedure is a {\em discretization} of the 
continuum theory, while the latter in addition provides a
{\em regularization}. In two-dimensional QRC no
cut-off related to geometry is needed \cite{localmeasure1}, 
while the DT-formalism 
explicitly introduces a lattice cut-off $\ep$. One can 
of course choose to introduce such a cut-off in two-dimensional 
QRC too, since it is needed in higher-dimensional QRC \cite{localmeasure2}.
However, it will make no difference for the arguments we present,
and since $\ep$ should be taken to zero at the end of the calculation,
we have chosen to 
work directly with $\ep =0$ for two-dimensional QRC.

Numerical simulations
have questioned if QRC agrees with continuum predictions
\cite{janke}.

\vspace{12pt}
\noindent
{\sl We will show that QRC is unlikely to be useful in 
quantum gravity. More precisely, we show that it does not 
reproduce the continuum theory of quantum gravity associated 
with \rf{genus}. It is difficult to say something rigorous about 
four-dimensional quantum gravity, since we have no well defined 
continuum theory with which we can compare. 
But the problems encountered in two-dimensional 
QRC are unchanged in higher dimensional QRC.}

\vspace{12pt}

There is presently no unique choice of measure to be used in QRC, in analogy 
with the DeWitt measure \rf{measure} in the continuum. {\em A priori}
this should not be viewed as a problem. Also in the case of DT 
there is no unique choice, but {\em universality} in the sense introduced 
in the theory of critical phenomena, should ensure that any reasonable 
change of measure is irrelevant in the continuum limit. This has been verified
in the context of DT by numerous numerical simulations.
In the next section we discuss the QRC-measures  proposed in the literature. 
In section 3 we analyze the functional integrals 
which result from these measures and show that none of the measures
will allow us to reproduce the continuum two-dimensional quantum gravity
theory. Section 4 contains our conclusions.

\section{Regge Integration Measures}

\subsection{DeWitt-like measure}
\label{sec:measure-disc}

We can repeat the construction \rf{line-element} of the 
norm of metric deformations
under the restriction that the deformations are constrained to 
represent the metric deformations allowed by QRC. For  
alternative discussions, starting from the canonical approach to gravity,
we refer to \cite{hamber} where the corresponding metric is 
denoted as the {\em Lund-Regge metric}, refering to an unpublished 
work of Lund and Regge. Since our derivation is valid in any 
dimension and has as its starting point the functional formalism,
we find it more appropriate to call the measure {\em DeWitt-like}.
  
First note that any $D$-simplex in a $D$-dimensional piecewise 
linear manifold $M$ can be covered with charts $(U,\phi)$, where
\begin{equation}
U=\{\xi\in 
{R}^D_{+}\,|\,\xi_1+\ldots+\xi_D<1\}
\end{equation}
and $\phi:U\rightarrow M$ is given by
\begin{equation}
\phi(\xi)=\xi_1y_1+\ldots+\xi_Dy_D+(1-\xi_1-\ldots-\xi_D)y_{D+1}\, ,
\end{equation}
where $y_1,\ldots,y_{D+1}$ are the vertices of the $D$-simplex.
In this way $M$ is equipped with a manifold structure. 
On this manifold we use a ``canonical'' 
metric, which on a $D$-simplex with chart $(U,\phi)$ is given by:
\begin{equation}
g_{\mu\nu}(\xi)= \frac{\partial\phi}{\partial\xi^{\mu}}\cdot
\frac{\partial\phi}{\partial\xi^{\nu}}
\label{canonical}
\end{equation}
This metric is compatible with the manifold structure
and has the advantage that it can be expressed solely in terms of the 
link-lengths of the piecewise linear manifold. 
In two dimensions \rf{canonical} has  the following form 
on a triangle
\begin{equation}
  g_{\mu\nu}=\left( \begin{array}{cc} x_1 & \frac{1}{2}(x_1+x_2-x_3)
  \\ \frac{1}{2}(x_1+x_2-x_3) & x_2 \end{array} \right)\, ,
\label{metric}
\end{equation}
where $x_i$ equals $l_i^2$, and $l_i$ are the link-lengths of the
triangle. We assume here and everywhere below that the link lengths $l_i$ 
satisfy the triangle inequalities.

The area $A$ of the triangle can be
expressed as
\begin{equation}
  A=\int
  d^2\xi\sqrt{g(\xi)}=\frac{1}{2}\sqrt{g}=\frac{1}{2}
  \big( x_1x_2-\frac{1}{4}(x_1+x_2-x_3)^2 \big)^{1/2}\, .
\end{equation}
The fluctuation of the metric given in terms of the $\delta x_i$'s is:
\begin{equation}
  \delta g_{\mu\nu}=\left( \begin{array}{cc} \delta x_1 & \frac{1}{2}(\delta
  x_1+\delta x_2-\delta x_3)
  \\ \frac{1}{2}(\delta x_1+\delta x_2-\delta x_3) & 
\delta x_2 \end{array} \right)\, .
\label{fluctuation}
\end{equation}
For a single triangle, the line element (\ref{line-element}) can now
be computed:
\begin{eqnarray}
\label{one-triangle}
|| \del g_{\m\n}||^2 & = & \int d^2\xi \sqrt{g(\xi)}\,\delta
g_{\mu\nu}(\xi)G^{\mu\nu\alpha\beta}\delta g_{\alpha\beta}(\xi)
\nonumber \\
& = & \frac{A}{2}\left(2\delta g_{\mu\nu} g^{\mu\alpha}g^{\nu\beta}\delta
  g_{\alpha\beta}+ C\delta g_{\mu\nu}g^{\mu\nu}g^{\alpha\beta} 
\delta g_{\alpha\beta}\right) \nonumber\\
& = & -2A\det \delta g_{\mu\nu}\det g^{\mu\nu}~~~~~~~~~~~~~~(C=-2) .
\end{eqnarray}
In the last line we have chosen the canonical value $-2$ for
$C$, which greatly simplifies the resulting expressions.
From \rf{metric} and \rf{fluctuation} we get
\begin{eqnarray}
|| \del g_{\m\n}||^2 & = & 
\frac{1}{16A}\left((\delta x_1)^2+(\delta x_2)^2+(\delta
  x_3)^2-2\delta x_1\delta x_2-2\delta x_1\delta x_3-2\delta x_2\delta
  x_3\right)  \nonumber \\
& = & [\,\delta x_1, \delta x_2, \delta x_3\,]\:\frac{1}{16A}\!\left[
  \begin{array}{rrr} 1 & -1 & -1 \\ -1 & 1 & -1 \\ -1 & -1 & 1
  \end{array} \right] \left[ \begin{array}{c} \delta x_1 \\ \delta x_2
  \\ \delta x_3 \end{array} \right] \, .
\end{eqnarray}
For a general two-dimensional triangulation with $L$ links the line
element is given by the sum of the line-elements 
(\ref{one-triangle}) over all triangles:
\begin{eqnarray}
  \label{general}
 ||\del g_{\m\n}||^2 & = & \sum_T\int d^2\xi \sqrt{g^T(\xi)}\,\delta
g^T_{\mu\nu}(\xi)(G^T)^{\mu\nu\alpha\beta}\delta g^T_{\alpha\beta}(\xi)
\nonumber \\
& = & [\,\delta x_1,\ldots,\delta x_L\,]\:{\bf M}\left[
  \begin{array}{c} \delta x_1 \\ \vdots \\ \delta x_L \end{array} \right]\, .
\end{eqnarray}
${\bf M}$ is a $L\times L$ matrix, which has the following structure:
\begin{figure}[htb]
  \begin{center}
\mbox{
\epsfxsize=6cm
\epsffile{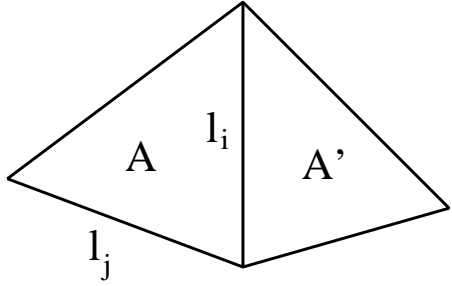}
}
\caption{Figure for explanation of the general structure of the matrix $M$.
  \label{fig:matrix}}
\end{center}
\end{figure}
For a closed surface each link appears in two triangles in the
sum (\ref{general}). With the notation depicted in figure
\ref{fig:matrix} we thus have the diagonal entry
\begin{equation}
  \label{madia}
  M_{ii} = \frac{1}{A} + \frac{1}{A'}\, , 
\end{equation}
and the off-diagonal entry
\begin{equation}
  \label{maoff}
  M_{ij} = M_{ji} = -\frac{1}{A}.
\end{equation}
Off-diagonal entries $M_{ij}$ equal zero if $l_i$ and $l_j$ are not
sides of the same triangle.
It follows that each row and column of $M$ has four non-vanishing 
off-diagonal entries for a closed surface.
Thus the matrix {\bf M} is a weighted adjacency matrix 
for the $\phi^4$-graph which is 
constructed by connecting the midpoints of neighboring links in the 
triangulation, with the weights given by \rf{madia} and \rf{maoff}.

Let $T$ denote the number of triangles.
From each row a factor
$(AA')^{-1}$ can be factorized. Since
each triangle has three sides, a factor $\prod_{k=1}^T
A_k^{-3}$ can be factorized from the determinant of {\bf M}:
\begin{eqnarray}
  \label{determinant}
  \det {\bf M} &=& \prod_{k=1}^T A_k^{-3}
  \left\vert
    \begin{array}{ccccccc}
      A_1+A_2 & -A_2 & -A_2 & -A_1 & -A_1 & 0 & \ldots \\
      \mc{7}{c}{\ldots}
    \end{array}
    \right\vert \nonumber \\
    &=:&  P(A_1,\ldots,A_T) \prod_{k=1}^T A_k^{-3}\, ,
\end{eqnarray}
where $P(A_1 , \ldots ,A_T)$ is a polynomial in the areas of the
triangles of the surface. $P$ vanishes whenever the area of 
two adjacent triangles
vanish, for example when one link goes to zero. This means that $P$ is
a highly non-local function, because each monomial of $P$ contains at
least half of the areas of the surface.

From \eref{general} we get that the DeWitt-like integration measure for
QRC is given by the square root of the determinant of $M$:
\begin{equation}
  \label{measure-disc}
  d\mu(l_1,\ldots,l_L) = \text{const}\times
  \frac{\sqrt{P(A_1 , \ldots ,A_T)}}{\prod_{k=1}^T A_k^{3/2}}
  \prod_{j=1}^L l_jdl_j\; \del ( \triangle)\, ,
\end{equation}
where $\del(\triangle)$ is a shorthand notation for the 
triangle inequalities satisfied by the links $l_i$.

We have not been able to obtain a closed form of $P$, but $P$ can be
determined in a number of special cases, as will be discussed below.
In the next section we will only need 
the fact that $P(A_i)$  is  
a function of the areas of the triangles and does 
not depend explicitly of the link length. 
 
The continuum measure (\ref{measure}) in two dimensions is
\begin{equation}
  [{\cal D}g_{\mu\nu}] = {\rm const}\times\prod_x g(x)^{-3/4}\prod_{\mu\leq\nu}
  dg_{\mu\nu}(x)
\end{equation}
Using $g(x)\sim A^2$, we see that the powers of the ``local'' areas in the
discretized measure coincide with the continuum measure. On the other hand
the discretized measure is non-local, whereas the continuum measure is
local. 

Below we will analyze certain properties of the measure 
\rf{measure-disc} and the analysis can be performed without further 
complications for the following generalization:
\begin{equation}
  \label{other-measures3}
  d\mu(l_1,\ldots,l_L) = \text{const}\times
  \frac{\sqrt{P(A_1^{2\b/3} , \ldots ,A_T^{2\b/3})}}{\prod_{k=1}^T A_k^{\beta}}
  \prod_{j=1}^L l_jdl_j \; \del ( \triangle)\, ,
\end{equation}
which appears when the measure \rf{one-triangle} is multiplied by the 
``local'' area factor $A_i^{1-\frac{2\b}{3}}$.  
This generalization highlights that QRC does not fix the powers 
of $A_i$ by any obvious principle, since the areas
are reparameterization invariant  objects which have 
no simple local continuum interpretation.

\subsection{Other Regge Measures}

It is clear that the measure \rf{measure-disc} is not  suited 
for numerical simulations since it is highly non-local. In addition
it should be clear that this measure is not forced upon us by 
the requirement of reparameterization invariance and ultra-locality 
in the same way as the continuum  DeWitt measure. It is not even 
the DeWitt measure restricted to the class of piecewise
linear metrics. This measure was constructed  in \cite{menotti},
as mentioned in the introduction. It is simply a ``translation'' of 
the DeWitt measure to the rather special class of piecewise linear 
manifolds used in QRC. Thus it is natural to ask if there are other 
measures which are local and which are equally good.
The most common local measures which have been used have the form 
(see \cite{localmeasure} and references therein)
\begin{equation}
  \label{other-measures1}
  d\mu(l_1,\ldots,l_L) = \prod_{j=1}^L \frac{dl_j}{l_j^{\alpha}}\; 
\del ( \triangle)
\end{equation}
and
\begin{equation}
  \label{other-measures2}
  d\mu(l_1,\ldots,l_L) = \frac{\prod_{j=1}^L
    l_jdl_j}{\prod_{k=1}^T A_k^{\beta}} \; \del ( \triangle)\, .
\end{equation}
The last measure is similar to the DeWitt measure except that the 
non-local term is missing.

\begin{figure}[htb]
  \begin{center}
\mbox{
\epsfxsize=11cm
\epsffile{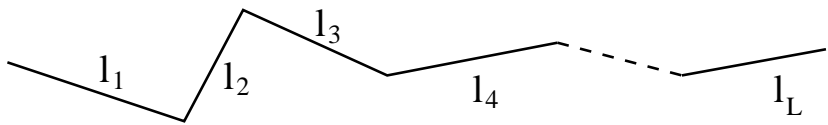}
}
\caption{A one-dimensional Regge-manifold\label{fig:onedim}}
\end{center}
\end{figure}

One argument for such  choices comes from the discretized Regge measure 
in one dimension. A one-dimensional Regge-manifold consists of 
$L$ straight lines of length $l_i$ ($x_i = l^2_i$), $i=1,\ldots,L$ 
(see fig.\ \ref{fig:onedim}). The ``canonical metric'' is thus
$g^i_{\mu\nu} = [ x_i]$. Using this, we
can immediately write down the norm:
\begin{equation}\label{one}
|| \del g_{\m\n} ||^2= 
\sum_{i=1}^L\int d\xi \sqrt{g^i(\xi)}\delta g^i(\xi)G^i\delta g^i(\xi)
= \frac{1}{2}(2+C)\sum_i x_i^{-3/2}(\delta x_i)^2
\end{equation}
The measure is thus given by
\begin{equation}
d\mu(l_1,\ldots,l_L)={\rm const}\times\prod_{i=1}^L x_i^{-3/4}dx_i=
{\rm const}\times\prod_{i=1}^L l_i^{-1/2}dl_i
\end{equation} 
In this case we have a a general factorization of 
$(1+C/2D)$ as in the continuum and the measure is local\footnote{We 
remark that precisely in $D=1$ one cannot use the canonical value 
$C=-2$. From the point of view of the continuum path integral this is 
not important since one can just choose a different value of $C$.}.
However, this is not true in higher dimensions, as shown above. One 
of the purposes of the derivation of the DeWitt-like measure in $D=2$ is 
to show that one cannot use \rf{one} as a motivation for \rf{other-measures1}
and \rf{other-measures2} in higher dimensions.
Nevertheless, from the point of view of universality all measures discussed 
above should be equally good, provided they lead to the correct 
continuum limit.

\subsection{The DeWitt-like Measure for Special Geometries}

In this sub-section we derive the DeWitt-like measure  for a number
of simple two-dimensional geometries.

\subsubsection{Two triangles glued together}
For the case of two triangles of area $A$ glued together along the links, the
total number of links $L$ is $3$, and the matrix $M$ assumes the following
form:
\begin{equation}
 M= \frac{1}{8A}\left[ \begin{array}{rrr} 1 & -1 & -1 \\ -1 & 1 & -1 \\
      -1 & -1 & 1 \end{array} \right]
\end{equation}
The measure $d\mu(l_1,l_2,l_3)$ is then
\begin{equation}
\label{blah}
  d\mu(l_1,l_2,l_3)= (\det M)^{1/2}\prod_{j=1}^3 l_jdl_j \; \del ( \triangle)=
  {\rm const}\times \frac{1}{A^{3/2}}\prod_{j=1}^3 l_jdl_j \; 
\del ( \triangle)\, .
\end{equation}

\subsubsection{Tetrahedron}
\begin{figure}[htb]
  \begin{center}
    \leavevmode
    \epsfig{file=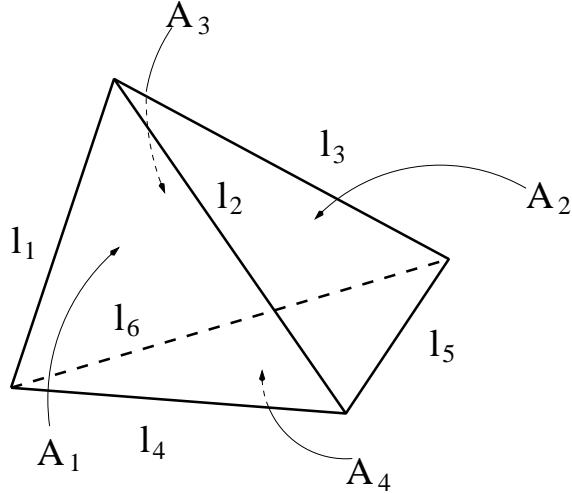, width=0.5\linewidth}
    \caption{Conventions for the tetrahedron.}
    \label{fig:tetra}
  \end{center}
\end{figure}

For the tetrahedron, the number $L$ of links is $6$, and the number $T$ of
triangles is $4$, see figure \ref{fig:tetra} for a parameterization.
With this parameterization, the matrix $M$ assumes the following form:
\renewcommand{\arraystretch}{1.5}
$$
  M=\frac{1}{16}\left[ \begin{array}{cccccc} \frac{1}{A_1}+\frac{1}{A_3} &
      -\frac{1}{A_1} & -\frac{1}{A_3} & -\frac{1}{A_1} & \quad\! 0 & -\frac{1}{A_3}
      \\ -\frac{1}{A_1} & \frac{1}{A_1}+\frac{1}{A_2} & -\frac{1}{A_2}
      & -\frac{1}{A_1} & -\frac{1}{A_2} & \quad\! 0 \\
      -\frac{1}{A_3} & -\frac{1}{A_2} & \frac{1}{A_2}+\frac{1}{A_3} & \quad\! 0 &
      -\frac{1}{A_2} & -\frac{1}{A_3} \\
      -\frac{1}{A_1} & -\frac{1}{A_1} & \quad\! 0 & \frac{1}{A_1}+\frac{1}{A_4} &
      -\frac{1}{A_4} & -\frac{1}{A_4} \\
      \quad\! 0 & -\frac{1}{A_2} & -\frac{1}{A_2} & -\frac{1}{A_4} & \frac{1}{A_2}
      +\frac{1}{A_4} & -\frac{1}{A_4} \\
      -\frac{1}{A_3} & \quad\! 0 & -\frac{1}{A_3} & -\frac{1}{A_4} & -\frac{1}{A_4}
      & \frac{1}{A_3}+\frac{1}{A_4} 
      \end{array} \right]
$$    
\renewcommand{\arraystretch}{1.0}
The determinant of this matrix can be computed, and one obtains the
result
\begin{equation}
  \det M = {\rm const}\times
  \frac{(\sum_{i=1}^4 A_i)^2}{\prod_{i=1}^4 A_i^2}\, .
\end{equation}
The measure is therefore given by
\begin{equation}
  d\mu(l_1,\ldots,l_6)={\rm const}\times\frac{\sum_{i=1}^4 A_i}{\prod_{i=1}^4 
A_i}  \;\prod_{j=1}^6 l_jdl_j \; \del ( \triangle)\, .
\end{equation}
Unfortunately the simple beauty of this formula does not extend to
more complicated geometries.

\subsubsection{$1\times n$-torus}
\begin{figure}[htbp]
  \begin{center}
    \leavevmode
    \epsfig{file=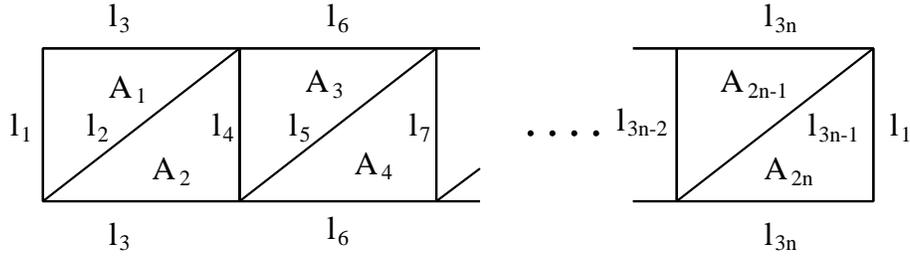,width=0.8\linewidth}
    \caption{Conventions for the $1\times n$-torus.}
    \label{fig:torus}
  \end{center}
\end{figure}
It is also possible to compute the measure for a thin torus explicitly.
The $1\times n$-torus is made of $3n$ links and $2n$ areas, see figure
\ref{fig:torus}.
\begin{figure}[htbp]
  \begin{center}
    \leavevmode
    \epsfig{file=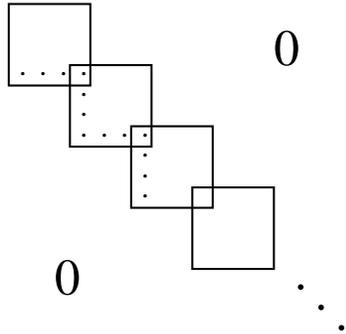,width=0.3\linewidth}
    \caption{Block-diagonal form of the matrix $M$.}
    \label{fig:block}
  \end{center}
\end{figure}
Because the links $l_{3k-1}$ and $l_{3k}$ are only
coupled to links inside one block of two triangles, the $3n\times 3n$-matrix
$M$ has a block-diagonal form, see figure \ref{fig:block}.
Block number $k$ has the following form:
\renewcommand{\arraystretch}{1.5}
\begin{equation}
  \label{one-block}
  \left[ \begin{array}{cccc} \frac{1}{A_{2(k-1)}}+\frac{1}{A_{2k-1}} &
      -\frac{1}{A_{2k-1}} & -\frac{1}{A_{2k-1}} & \quad\! 0 \\
      -\frac{1}{A_{2k-1}} & \frac{1}{A_{2k-1}}+\frac{1}{A_{2k}} &
      -\frac{1}{A_{2k-1}}-\frac{1}{A_{2k}} & -\frac{1}{A_{2k}} \\
      -\frac{1}{A_{2k-1}} & -\frac{1}{A_{2k-1}}-\frac{1}{A_{2k}} &
      \frac{1}{A_{2k-1}}+\frac{1}{A_{2k}} & -\frac{1}{A_{2k}} \\
      \quad\! 0 & -\frac{1}{A_{2k}} & -\frac{1}{A_{2k}} &
      \frac{1}{A_{2k}}+\frac{1}{A_{2k+1}} \end{array} \right] \, .
\end{equation}
\renewcommand{\arraystretch}{1.0}
By making simple row- and column-transformations on the matrix $M$
one can extract a factor
$4^n \prod_{k=1}^n \left(\frac{1}{A_{2k-1}}+\frac{1}{A_{2k}}\right)$
from the determinant of $M$. We expand the remaining determinant using
its linearity in the rows and in the columns. The result is
\begin{equation}
  \det M = (-4)^n\prod_{k=1}^n \left(\frac{1}{A_{2k-1}}+\frac{1}{A_{2k}}\right)
  \cdot\left(\prod_{k=1}^n \frac{1}{A_{2k-1}} + (-1)^{n-1}\prod_{k=1}^n
    \frac{1}{A_{2k}}\right)^2\, .
\end{equation}
Thus the DeWitt-measure for the $1\times n$-torus is
\begin{eqnarray}
  \label{thin-torus}
  d\mu(l_1,\ldots,l_{3n}) & = & {\rm const}\times\frac{\prod_{j=1}^{3n} l_j
    dl_j \; \del ( \triangle)}{\prod_{k=1}^{2n}A_k^{3/2}}\prod_{k=1}^n
\left(A_{2k-1}+A_{2k}
      \right)^{1/2}\cdot \nonumber \\
      & & \left|\,\prod_{k=1}^n A_{2k}+(-1)^{n-1}\prod_{k=1}^n
        A_{2k-1}\,\right| \, .
\end{eqnarray}
Note that the measure of this special geometry might vanish if $n$ is
even. This happens for instance if all triangles have the same area.

\subsubsection{Hedgehog geometry}

\begin{figure}[htbp]
  \begin{center}
    \leavevmode
    \epsfig{file=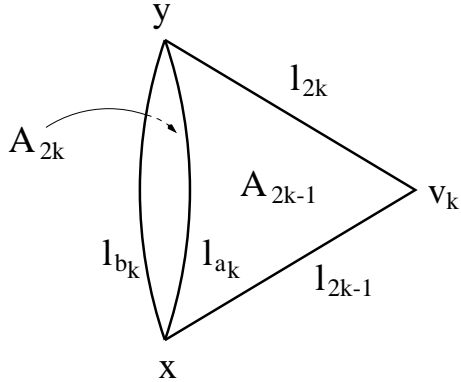,width=0.4\linewidth}
    \caption{Elements of the hedgehog.}
    \label{fig:hat}
  \end{center}
\end{figure}
As a last example we compute the DeWitt  measure
for a somewhat singular ``hedgehog'' geometry which 
has played an important role in the analysis of the 
DT-version of two-dimensional quantum gravity coupled to 
matter fields \cite{dt}. The building blocks of the hedgehog geometry  
are ``hats'' like the one shown in fig.\ \ref{fig:hat}. 
This element consists of two triangles which are glued
together along the links $l_{2k-1}$ and $l_{2k}$. The areas of the triangles
are $A_{2k-1}$ and $A_{2k}$, respectively. Now we take $N$ such elements
and glue them together successively along the links
$l_{b_k}$ and $l_{a_{k+1}}$ to make a hedgehog-like geometry.
After the gluing the vertices $x$ and $y$ will have the order $2N$, whereas
the vertices $v_k$ have the order two.

If we now write down the rows and columns of $M$ in the order
$l_{a_k}$, $l_{2k-1}$, $l_{2k}$, $l_{b_k}, \ldots$, we see that $M$
takes indeed the same form as the matrix $M$ for the thin torus
in the last section. Therefore the measure will also be given by
equation (\ref{thin-torus}).

\section{Diseases of Quantum Regge Calculus}

Let us recall some of the results from continuum 2d Quantum gravity
based to the functional integral \rf{*1} in the case of manifolds
with no handles\footnote{Since all critical indices which can be calculated 
in continuum quantum gravity, using either the KPZ or the DDK formalism, 
agrees with the critical indices which can be calculated using the 
DT-formalism, we consider the two as equivalent. There 
are a number of results  which can only be obtained using the DT-formalism.
We will still denote these ``continuum results''.}. The partition function is 
\begin{equation}
\label{Z}
Z(\L) = \L^{-5/2},
\end{equation}
and the Hartle-Hawking wave-functional for a universe with boundary 
which has length $\ell$ and one marked point is given by  
\begin{equation}
\label{W}
W(\ell,\L) = \frac{1}{\ell^{5/2}}\Bigl( 1 + \ell\sqrt{\L}\Bigr)\, 
\e^{-\ell \sqrt{\L}}  . 
\end{equation}
Finally, let us consider (closed) universes of fixed volume $V$ (i.e.\ fixed 
area since we consider two-dimensional universes). The 
(suitable normalized) partition function
for such universes, where in addition two marked points are separated 
a geodesic distance $R$, is given by 
\begin{equation}
\label{twopt}
Z(R,V) = V^{-1/4} F(R/V^{\oq}),
\end{equation}   
where $F(x)$ can be expressed in terms of certain hyper-geometric functions
\cite{aw,ajw}. Here we only need the fact that $F(x)$ behaves as 
\begin{equation}
\label{largex}
F(x) \sim \e^{-x^{4/3}}~~~~~{\rm for}~~~~~x \to \infty.
\end{equation}
From \rf{twopt} and \rf{largex} it follows that 
 {\em any} power of the average radius can be calculated as 
\begin{equation}
\label{fractal}
\la R^n \ra_V  = \int_0^\infty d\tilde{R}\; Z(\tilde{R},V)\, \tilde{R}^n
\sim V^{n/4}.
\end{equation}
This relation shows that the fractal dimension of 
two-dimensional quantum gravity is equal four ! \cite{japanese,aw}.

While we are unable to calculate \rf{Z} and \rf{W} in QRC, we can 
show that \rf{fractal} is not satisfied in QRC. In fact, we will prove
that in general \rf{fractal} is not defined in QRC:

\vspace{12pt}
\noindent
{\sl For any $\a$ or $\b$ in \rf{other-measures3}, \rf{other-measures1}
and \rf{other-measures2} there exists an $n$ such that for any link
$l$ in a given triangulation and any (fixed) value of the 
space-time volume
\begin{equation}
\label{infty}
\la l^n \ra_V = \infty.
\end{equation}
}
As a consequence of this theorem  the average radius $\la R \ra$, 
or some suitable power $\la R^n \ra$, does not exist.  This is in sharp 
contrast to this situation encountered in eq.\ \rf{fractal}. It shows
that the ensemble average over geometries defined in QRC 
has no genuine intrinsic scale set by the volume of space-time or 
equivalently by the 
cosmological constant. In this way the theory differs radically from 
the corresponding continuum quantum gravity theory.

The proof is as follows:
\begin{figure}[htbp]
  \begin{center}
    \leavevmode
    \epsfig{file=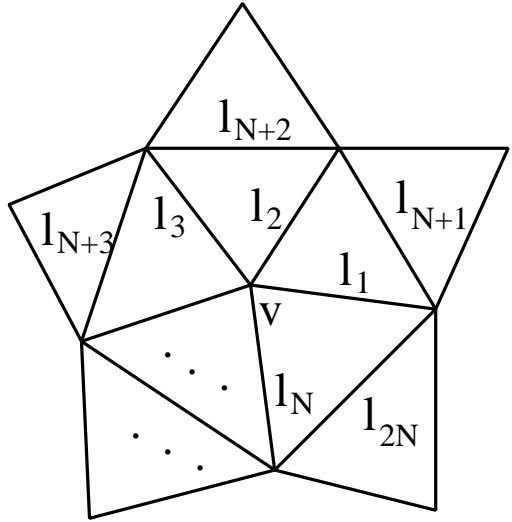, width=0.45\linewidth}
    \caption{Parameterization of the link lengths.}
    \label{fig:supplfig}
  \end{center}
\end{figure}
Let the link $l_1$ be connected to a vertex $v$ of coordination
number $N$, see figure \ref{fig:supplfig}. We want to analyze the
situation where the vertex
$v$ is pulled to infinity while the global area of the surface is held
bounded. This can be done by keeping the link-lengths
$l_{N+1}, \ldots, l_{2N}$ of the order $l_1^{-1}$.
Around $v$ the link length $l_1$ can be integrated freely from
a large length $L$ to infinity. Then the integration of $l_2, \ldots, l_N$
is constrained by triangle inequalities.
For the measure \eref{other-measures1} the integration over
$l_2, \ldots, l_N$ yields $(l_{N+2} \ldots l_{2N})^{\frac{N-1}{N}}
l_1^{(1-N)\alpha}$. Here we symmetrized over the links
$l_{N+1}, \ldots, l_{2N}$.
The functional
integral in the configuration space corresponding to this situation
around the vertex $v$ is then given by
\begin{equation}
  \label{sup1}
  \int_L^{\infty} dl_1 l_1^{-N \alpha} \int_0^{L/l_1} dl_{N+1}\ldots dl_{2N}
  (l_{N+1} \ldots l_{2N} )^{\frac{2N-1}{N}-\alpha}\, .
\end{equation}
The additional factor $l_{N+1} \ldots l_{2N}$ in \rf{sup1} originates from 
the triangle inequalities for the adjacent triangles. The integral over
$l_{N+1}, \ldots, l_{2N}$ exists if $\frac{2N-1}{N} - \alpha >-1$, i.e.\
\begin{equation}
  \label{sup2}
  \alpha < 3-\frac{1}{N}\, .
\end{equation}
Thus the $l_1$ integration becomes
\begin{equation}
  \label{sup3}
  \int_L^{\infty} dl_1 l_1^{-N\alpha} \left( \frac{L}{l_1}
  \right)^{N(\frac{3N-1}{N}-\a)}
  \propto
  \int_L^{\infty} dl_1 l_1^{1-3N}\, ,
\end{equation}
This shows that $\langle l_1^{3N-2} \rangle = \infty $ which completes
the proof for the measure \eref{other-measures1}.
In a similar way one can analyze the two other measures
\eref{other-measures2} and \eref{other-measures3}. To this end we
have to parameterize the areas in terms of the link lengths. In
this part of the configuration space the areas are given by the product
of a small link  length and a long link length leading to extra
factors $l_1^{-\beta} \ldots l_N^{-\beta}$ times
$l_{N+1}^{-2\beta} \ldots l_{2N}^{-2\beta}$. By performing  the same analysis as
above for \eref{other-measures2} 
we see that the functional integral is proportional to
\begin{equation}
  \label{sup4}
  \int_L^{\infty} dl_1 l_1^{1-3N+N\beta}\, .
\end{equation}
Thus $\langle l_1^n \rangle = \infty$ for $n \geq N(3-\beta)-2$.

For \eref{other-measures3} one can extract a factor of $\prod_{i=1}^L
l_i^{1/2}$ from the non-local factor $\sqrt{P}$. Performing the analysis
as above we get again the result \eref{sup4}.
Thus the theorem is proved.

In the following we will analyze \rf{infty} for a 
number of different triangulations in order to 
illustrate how sensitive the QRC-measures  are to the choice 
of triangulation, which is another worrisome aspect of this 
formalism. For some triangulations the measure itself is ill defined,
for some triangulations $\la l \ra = \infty$ while for other measures
the expectation value of some link $l$ may be finite while the expectation 
value of $l^2$ is infinite etc. We denote $\la l^n \ra = \infty$ 
as the {\em appearance of spikes}.

\subsection{Spikes in a hexagonal geometry}

Let us first analyze the case of a hexagonal triangulation.
\begin{figure}[htb]
  \begin{center}
    \leavevmode
    \epsfig{file=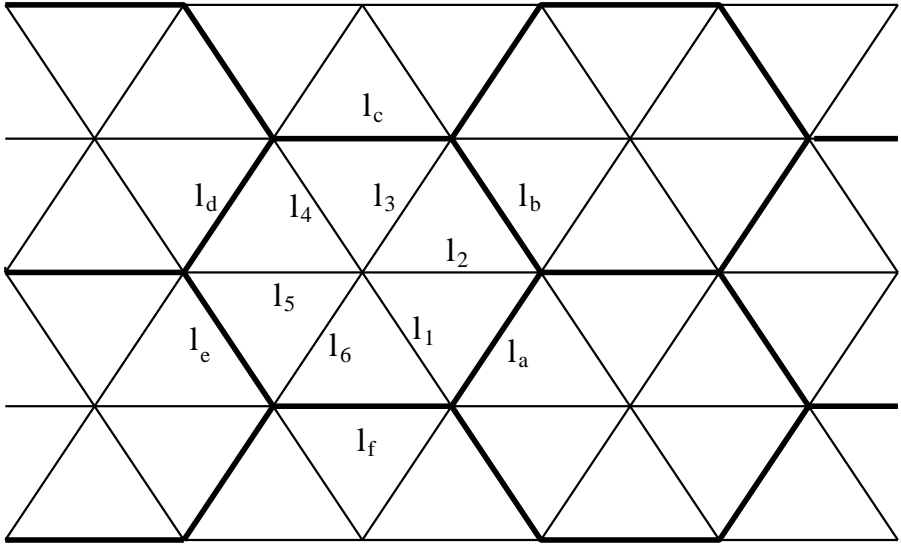,width=0.8\linewidth}
    \caption{Only spikes in a hexagonal geometry}
    \label{fig:hexa}
  \end{center}
\end{figure}
We subdivide a regular triangulation as in figure \ref{fig:hexa} into
$N$ hexagonal cells. We analyze the functional integral in that part
of the configuration space, where all centers of the cells form spikes.
This means that the $6N$ links $l_1$, $\ldots$, $l_6$,
etc.\ are very large.
Furthermore, because the action is proportional to the area of the
whole surface, we have to keep the area bounded to prevent exponential
damping. Therefore the $3N$ link lengths $l_a, \ldots , l_f$, etc.\ 
have to be very small, of the order of $l_1^{-1}$.

In cell number $k$ one link, $l_{6k+1}$, can be integrated freely from a
large length $L$ to $\infty$. The integration of $l_{6k+2},\ldots,l_{6k+6}$
is then constrained by triangle inequalities.
For the measure (\ref{other-measures1}) the integration over
$l_{6k+2},\ldots,l_{6k+6}$ thus yields a factor $(l_{a_k}\cdots
l_{f_k})^{\frac{5}{6}} l_{6k+1}^{-5\alpha}$ in the $k$'th cell. Here
we have symmetrized over all the links $l_{a_k},\ldots,l_{f_k}$
of the $k$'th cell. Noting that we can assign the
three links $l_{a_k}, l_{b_k}, l_{c_k}$ to the $k$'th cell,
the functional-integral in the spiky part of
the configuration space is proportional to:
\begin{eqnarray}
  \label{spikes}
  \int_{L}^{\infty} \prod_{k=1}^{N} l_{6k+1}^{-6\alpha} dl_{6k+1}
  \int_{0}^{L/l_{6k+1}} \prod_{k=1}^{N} \left( l_{a_k} l_{b_k} l_{c_k}
  \right)^{\frac{5}{3}-\alpha} dl_{a_k}dl_{b_k}dl_{c_k}\, .
\end{eqnarray}
The integration of the links $l_{a_k}, l_{b_k}, l_{c_k}$
only exists if
\begin{equation}
  \label{cond1}
  \alpha < \frac{8}{3}\, .
\end{equation}
If we perform the integration, an integral of the following 
form will remain:
\begin{equation}
  \label{rest-stuff}
  \int_{L}^{\infty} \prod_{k=1}^{N} l_{6k+1}^{-3\alpha-8} dl_{6k+1}\, ,
\end{equation}
which only exists if
\begin{equation}
  \label{cond2}
  \alpha > -\frac{7}{3}\, .
\end{equation}
Furthermore, from (\ref{rest-stuff}) we see that the expectation value
of $l_1^{n}$ will be infinite if $n$ is larger or equal $7+3\alpha$.

If we analyze the measure \eref{other-measures2} along the same lines, we
have to para\-me\-tri\-ze the areas of the triangles
in terms of the link lengths. For a spiky
geometry the areas are given by products of a small and a long link in the triangle. We thus get extra factors of
$l_{6k+1}^{-\beta}\cdots l_{6k+6}^{-\beta}$
and $l_{a_k}^{-2\beta} \cdots l_{f_k}^{-2\beta}$. 
Then we can perform the
analysis as above and get the bound
\begin{equation}
  \label{cond3}
  \beta < \frac{11}{6}
\end{equation}
for $\beta$. In this case we get no lower bound. The expectation value
of $l_1^{n}$ will be infinite if $n\geq 4$. 

For this geometry one can extract a factor $\prod_{i=1}^L l_i^{1/2}$
from the nonlocal factor $\sqrt{P}$ of the measure \eref{other-measures3}.
Consequently, we get for this measure  extra factors of
$l_{6k+1}^{\frac{1}{2}-\beta}\cdots l_{6k+6}^{\frac{1}{2}-\beta}$
and $l_{a_k}^{\frac{1}{2}-2\beta} \cdots l_{f_k}^{\frac{1}{2}-2\beta}$
from the areas. Using this we get the bound
\begin{equation}
  \label{cond4}
  \beta < \frac{25}{12}
\end{equation}
and the expectation value of $l_1^n$ will be infinite 
for $n\geq \frac{5}{2}$.

\subsection{Spikes in a 12-3 geometry}

\begin{figure}[htb]
  \begin{center}
    \leavevmode
    \epsfig{file=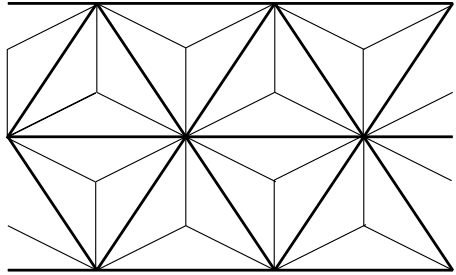,width=0.4\linewidth}
    \caption{ A 12-3 geometry. All vertices
    of order 3 form spikes.}
    \label{fig:12-3}
  \end{center}
\end{figure}
In a similar way we can analyze the partition function for a regular toroidal
triangulation where $N$ vertices have order 3 and $N/2$ vertices
have order 12, see figure \ref{fig:12-3}. We can form spikes by pulling the
vertices of order 3 to infinity keeping the area bounded from above.
The bounds for the measures \eref{other-measures1}, \eref{other-measures2} 
and \eref{other-measures3} are given in table \ref{tab:bounds1}.
Note that these bounds are sharper than for the hexagonal geometry.
\renewcommand{\arraystretch}{1.2}
\begin{table}[htb]
  \begin{center}
    \leavevmode
    \begin{tabular}[c]{|c|c|c|} \hline
      measure & bounds & values of $n$ s.t.
      $\langle l^n \rangle < \infty$ \\ \hline
      \eref{other-measures1} & $-\frac{5}{3}
      < \alpha < \frac{7}{3}$ & $n<\frac{5}{2} + 3\alpha$ \\
      \eref{other-measures2} & $\beta < \frac{5}{3}$ & $n<1$ \\
      \eref{other-measures3} & $\beta <
      \frac{23}{12}$ & $n<\frac{1}{4}$ \\ \hline
    \end{tabular}
    \caption{Bounds for the exponents $\alpha$ and $\beta$
      for the measures \eref{other-measures1},
      \eref{other-measures2} and \eref{other-measures3} and values
      of $n$ for which $\langle l^n\rangle$ is finite for the
      12-3 geometry.}
    \label{tab:bounds1}
  \end{center}
\end{table}
\renewcommand{\arraystretch}{1.0}

\subsection{Spikes in a hedgehog-like geometries}

\begin{figure}[htb]
  \begin{center}
    \leavevmode
    \epsfig{file=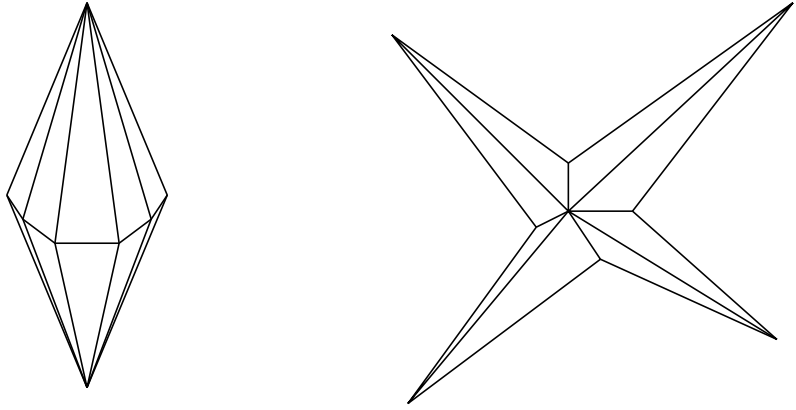,width=0.7\linewidth}
    \caption{A degenerate geometry. Spikes can be formed in several ways.}
    \label{fig:degenerate}
  \end{center}
\end{figure}
One can get sharper bounds for more degenerate geometries.  As explained
above, these kind of geometries occur in several phases of quantum
gravity in the context of DT. However, the considerations are 
essentially local and for very large triangulations it would 
be an unnatural constraint on the possible choices of triangulations
if hedgehog-like connectivity should not be allowed locally.
As a first example we analyze the geometry depicted 
in figure \ref{fig:degenerate}. Here
there are two vertices of order $N$ while all other vertices have order
4. This is a slightly more regular configuration than the hedgehog
geometry associated with figure \ref{fig:hat}.

If spikes are formed by every second vertex of order 4, see the
second part of figure \ref{fig:degenerate}, the bounds 
take the form given in upper part of table \ref{tab:bounds3}.

For a hedgehog configuration constructed from the spikes in figure 
\ref{fig:hat} we get the following bounds recorded in the lower part of 
table \ref{tab:bounds3}, when the spike-vertices
are allowed to go to infinity.

\renewcommand{\arraystretch}{1.2}
\begin{table}[htb]
  \begin{center}
    \leavevmode
    \begin{tabular}[c]{|c|c|c|} \hline
      measure & bounds & values of $n$ s.t.
      $\langle l^n \rangle < \infty$ \\ \hline
      \eref{other-measures1} & $-2
      < \alpha < \frac{5}{2}$ & $n<2\alpha+4$ \\
      \eref{other-measures2} & $\beta < \frac{7}{4}$ & $n<2$ \\
      \eref{other-measures3} & $\beta <
      2$ & $n<1$ \\ \hline \hline
       \eref{other-measures1} 
     & $-1  < \alpha < 2$ & $n<1+\alpha$ \\
      \eref{other-measures2} & $ \mbox{partition function undefined}$ 
& $n< 0$ \\
      \eref{other-measures3} & $\mbox{partition function undefined}$ & 
       $n<-\oh$ \\ \hline
    \end{tabular}
    \caption{The upper part of the table
shows bounds for the exponents $\alpha$ and $\beta$
      for the measures \eref{other-measures1},
      \eref{other-measures2} and \eref{other-measures3} and values
      of $n$ for which $\langle l^n\rangle$ is finite in the case
      illustrated in the second part of figure \ref{fig:degenerate}.
The lower part of table shows the similar bounds for hedgehog 
      geometries constructed from the building blocks
      illustrated in figure \ref{fig:hat}.  }
    \label{tab:bounds3}
  \end{center}
\end{table}
\renewcommand{\arraystretch}{1.0}

\subsection{One vanishing link}

Besides spikes one can also analyze the behavior of the functional
integral if one or more links go to zero as in
\begin{figure}[htb]
  \begin{center}
    \leavevmode
    \epsfig{file=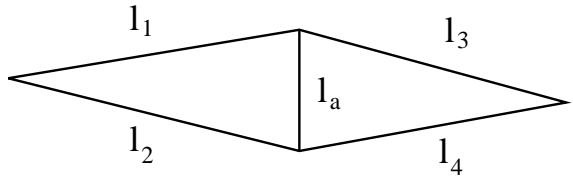,width=0.5\linewidth}
    \caption{One link length ($l_a$) vanishes, the other
      stay finite. The whole area is again bounded from above.}
    \label{fig:onelink}
  \end{center}
\end{figure}
figure \ref{fig:onelink}.
In the notation of figure \ref{fig:onelink}
we demand that $l_a$ is small whereas the other link length are
bounded away from zero and from infinity. Then one can integrate the
links $l_1$ and $l_3$ freely and  the integration of
$l_2$ and $l_4$ is constrained by triangle inequalities. Their integration
therefore yields a factor $l_a^2$ and some powers of $l_1$ and $l_3$
which depend on the measures.
The resulting bounds are given in table \ref{tab:bounds}.
\renewcommand{\arraystretch}{1.2}
\begin{table}[htb]
  \begin{center}
    \leavevmode
    \begin{tabular}[c]{|c|c|}\hline 
      \mc{1}{|c|}{measure} & 
      \mc{1}{c|}{bound on exponent} \\ \hline
      \eref{other-measures1} & $\alpha<3$ \\
      \eref{other-measures2} & $\beta<2$ \\
      \eref{other-measures3} & $\beta < \frac{9}{4}$ \\ \hline
    \end{tabular}
    \caption{Bounds on the exponents $\alpha$ and
      $\beta$ for the case of one vanishing link length.}
    \label{tab:bounds}
  \end{center}
\end{table}
\renewcommand{\arraystretch}{1.0}

\section{Conclusion}
\label{sec:conlusio}

We have shown that the predictions of Quantum Regge Calculus 
do not agree with known results of continuum two-dimensional 
quantum gravity, as defined by the functional \rf{*1}.
No natural scale seems to emerge from measures 
\rf{other-measures1}-\rf{other-measures3} and even expectation values 
of suitable  powers of the length of a {\em single} link will diverge.
The probability of having arbitrary large link-length will never 
be exponentially suppressed even if the volume of space-time is kept 
bounded. This is in contrast to the situation for continuum two-dimensional 
quantum gravity where the probability to have two points 
separated a distance $R$ is exponentially suppressed as 
$$
\exp (-R\sqrt{\L})
$$
for given cosmological constant, and suppressed as 
$$
\exp \left\{-\Bigl(\frac{R}{V^{\oq}}\Bigr)^{4/3}\right\}
$$ 
if the volume of space-time is kept fixed to be $V$ \cite{aw}.

In addition  the finiteness of $\la l^n \ra$ for given 
values of $\a$ and $\b$ (and small $n$ where the expectation value has 
a change to exist) depends crucially on the chosen fixed triangulation
in QRC. In fact, for given value of $\a$ and $\b$ the bare existence 
of the partition function could depend on the chosen triangulation.
Clearly this is not a desirable situation.

We expect the situation to be similar in higher dimensional 
QRC. The measures \rf{other-measures1} and \rf{other-measures2}
have an obvious generalization to higher dimensional simplicial
manifolds. The DeWitt-like measure for a $D$-dimensional simplex 
can be obtained from the norm
\begin{equation}\label{xstra1}
|| \del g_{\m\n}||^2_D = \del x \, {\bf M}_D \, \del x\, ,
\end{equation}
where $\del x$ is the $D(D+1)/2$ dimensional vector of $x_i=l_i^2$ deviations.
As in the two-dimensional case the line element 
for the $D$-dimensional triangulation
is the sum of line elements \rf{xstra1} for the individual simplexes 
constituting the triangulation. Thus the structure of the ${\bf M}_D^{total}$
is an obvious superposition of the basic matrices \rf{xstra1} in a way 
similar to the two-dimensional case (see \rf{general}). 

For instance, in $D=3$ the matrix ${\bf M}_D$ is 6-dimensional with 
entries depending on $x_i$'s.  The measure 
\begin{equation}\label{xstra2}
d\m = \sqrt{\det {\bf M}^{total}_D} \; \prod_{i=1}^L dx_i \; \del(\triangle)
\end{equation}
for the closed manifold, constructed from two tetrahedra 
glued together, is equal to
\begin{equation}\label{xstra3}
d\m (l_1,\dots,l_6) = \frac{\rm const.}{V} 
\prod_{i=1}^6 l_idl_i \; \del(\triangle),
\end{equation}   
where $V$ is the tree-volume of the manifold.
Thus the partition function of the two-simplex gravity is given by 
\begin{equation}\label{xstra4}
\int \e^{-\L V + \sum_{i=1}^6 l_i(2\pi  -2 \a_i)} \; \del(\triangle)\;
\frac{1}{V}\prod_{i=1}^6 l_idl_i
\end{equation}
where $\a_i$ are the dihedral angles. \rf{xstra3} is the three-dimensional 
expression corresponding to the \rf{blah}. Already the partition function
corresponding to \rf{blah} is 
divergent (see table \ref{tab:bounds3}) 
and producing spikes and \rf{xstra4} is even less well
defined due to unboundedness of the action in three dimensions. 
It is well known that this disease has to be cured by a 
cut-off involving for instance higher derivative terms (for a definition
in the context of integral geometry, see \cite{as}). However, the 
problem with spikes will reappear as the cut-off is taken 
to zero and even if we complete drop the curvature term 
from the action the  partition function will not be well defined,
precisely as in the two-dimensional case.
Although we have not investigated higher dimensional cases in detail it seems 
hard to believe that QRC is a viable candidate for the 
quantum gravity when it is unable to reproduce the results 
of the simplest known such theory, {\em viz.} two-dimensional 
quantum gravity.

\vspace{24pt}

\subsection*{Acknowledgment}
J.A. acknowledges the support of the Professor Visitante Iberdrola
grant and the hospitality at the University of Barcelona, where part
of this work was done.
J.R. gratefully acknowledges financial support by the Studienstiftung
des deutschen Volkes. G.K.S. thanks NBI for warm hospitality.

\end{document}